\documentclass[a4paper,10pt,runningheads]{llncs}
\usepackage{hyperref}
\usepackage{includes}

\begin{document}

\title{Computer certified efficient exact reals in \Coq}
\author{Robbert Krebbers \and Bas Spitters\thanks{The research leading to these results has received funding from the European Union's 7th Framework Programme under grant agreement nr.~243847 (ForMath).}}
\institute{Radboud University Nijmegen}
\maketitle

\begin{abstract}
Floating point operations are fast, but require continuous effort on the part of the user in order to ensure that the results are correct.
This burden can be shifted away from the user by providing a library of \emph{exact} analysis in which the computer handles the error estimates.
We provide an implementation of the exact real numbers in the \Coq{} proof assistant. This improves on the earlier \Coq-implementation by O'Connor in two ways: we use dyadic rationals built from the machine integers and we optimize computation of power series by using approximate division. Moreover, we use type classes for clean mathematical interfaces. This appears to be the first time that type classes are used in heavy computation. We obtain over a 100 times speed up of the basic operations and indications for improving the \Coq\ system.
\end{abstract}

\section{Introduction}
Real numbers cannot be represented exactly in a computer. Hence, in constructive analysis~\cite{Bishop67} one approximates real numbers by rational, or dyadic numbers. The real numbers are the completion of the rationals. This completion construction can be organized in a monad, a familiar construct from functional programming (Section~\ref{section:completion-monad}). The completion monad provides an efficient combination of proving and computing~\cite{OConnor:mscs}. In this way, O'Connor~\cite{Oconnor:real} implements exact real numbers and the transcendental functions on them in \Coq.

A number of possible improvements in this implementation were already suggested in~\cite{Riemann}. First, we can use \Coq's new machine integers; see Section~\ref{section:machine}. Second, we can use dyadic rationals (that are numbers of the shape $n * 2 ^ e$ for $n,e \in \Z$, also known as infinitary floats). Third, the implementation of power series can be improved by using approximate division. Here we carry out all three optimizations. Unfortunately, changing O'Connor's implementation to use the new machine integers was far from trivial, as he used a particular concrete representation of the rationals. To avoid this in the future, we provide an abstract specification of the dense set as \emph{approximate rationals}; see Section~\ref{section:interfaces}.

In Section~\ref{section:interfaces} we provide some abstract order theory culminating in the theory of approximate rationals. Section~\ref{section:series} deals with computing power series using dyadics. Section~\ref{section:Wolfram} describes Wolfram's algorithm to compute the square root of a real number. We finish with some benchmarks in Section~\ref{section:bench}.

\section{The \Coq-system}
The \Coq{} proof assistant is based on the calculus of inductive constructions~\cite{CoquandHuet,CoquandPaulin}, a dependent type theory with (co)inductive types; see~\cite{Coq,BC04}. In true Curry-Howard fashion, it is both a pure functional programming language with an expressive type system, and a language for mathematical statements and proofs. We highlight some aspects of \Coq{} relevant for our development.

\paragraph{Types and propositions.}
Propositions in \Coq{} are types~\cite{ITT,MartinLof:1982}, which themselves have types called \emph{sorts}. \Coq{} features a distinguished sort called \prop\ that one may choose to use as the sort for types representing propositions. The distinguishing feature of the \prop\ sort is that terms of non-\prop\ type may not depend on the values of inhabitants of \prop\ types (that is, proof terms). This regime of discrimination establishes a weak form of proof irrelevance, in that changing a proof can never affect the result of value computations. On a practical level, this lets \Coq{} safely erase all \prop\ components when extracting certified programs to \OCaml{} or \Haskell. We should note however, that in practice, \Coq's extraction mechanism~\cite{letouzey2008extraction} is still very hard to use for programs with the complexity, in terms of depth of definitions, that we are interested in \cite{cruzfilipe2003,cruzfilipe2006}.

\paragraph{Equality, setoids, and rewriting}
Because the \Coq{} type theory lacks quotient types (as it would make type checking undecidable), one usually bases abstract structures on a \emph{setoid} (`Bishop set'): a type equipped with an equivalence relation~\cite{Bishop67,Hofmann}. This leads to a naive set theory as described by Palmgren~\cite{palmgren2009constructivist}. When the user attempts to substitute a given (sub)term using an equality, the system keeps track of, resolves, and combines proofs of equivalence~\cite{Setoid-rewrite}.

The `native' notion of equality in \Coq{}, \emph{Leibniz equality}, is that of terms being convertible, naturally reified as a proposition by the inductive type family \lstinline|eq| with single constructor \lstinline|eq_refl : ∀ (T : Type) (x : T), x ≡ x|, where \lstinline|a ≡ b| is notation for \mbox{\lstinline|eq T a b|}. Since convertibility is a congruence, a proof of \lstinline|a ≡ b| lets us substitute \lstinline|b| for \lstinline|a| anywhere inside a term without further conditions. Our interest is in more complicated equalities, so we diverge from \Coq{} tradition and reserve \lstinline|=| for setoid equality.
Rewriting with \lstinline|=| \emph{does} give rise to side conditions. For instance, consider formal fractions of integers as a representation of rationals. Rewriting a subterm using such an equality is permitted only if the subterm is an argument of a function that has been proven to \emph{respect} the equality. Such a function is called \lstinline|Proper|, and that property must be proved for each function in whose arguments we wish to enable rewriting.

\paragraph{Type classes.}
Type classes have been a great success story in the \Haskell{} functional programming language, as a means of organizing interfaces of abstract structures. \Coq's type classes provide a superset of their functionality, but are implemented in a different way.

In \Haskell{} and \Isabelle, type classes and their instances are second class. They are handled as specialized syntactic constructs whose semantics are given specifically by the type class apparatus. By contrast, the expressivity of dependent types and inductive families as supported in \Coq, combined with the use of pre-existing technology in the system (namely proof search and implicit arguments) enable a \emph{first class} type class implementation~\cite{DBLP:conf/tphol/SozeauO08}: classes are ordinary record types (`dictionaries'), instances are ordinary constants of these record types (registered as \emph{hints} with the proof search machinery), class constraints are ordinary implicit parameters, and instance resolution is achieved by augmenting the unification algorithm to invoke ordinary proof search for implicit arguments of class type.
Thus, type classes in \Coq{} are realized by relatively minor syntactic aids that bring together existing facilities of the theory and the system into a coherent idiom, rather than by introduction of a new category of qualitatively different definitions with their own dedicated semantics.

We use the algebraic hierarchy based on type classes and its abstract specification of $\N,\Z$ and $\Q$ described in \cite{math-classes}. Unfortunately, we should note that we have clearly met the efficiency problems connected to the current implementation of type classes in \Coq. Luckily, these efficiency problems are limited to instance resolution which is only performed at compile time. Type classes have only a very minor effect on the computation time of type checked terms due to the absence of code inlining; see Section~\ref{section:bench} for timings. 

\paragraph{Virtual machine and machine integers.}
\label{section:machine}
\Coq{} includes a virtual machine~\cite{Compiler}, \lstinline|vm_compute,| based on \OCaml{}'s virtual machine to allow efficient evaluation. 
Both the abstract machine and its compilation scheme have been proved correct, in \Coq, with respect to the weak reduction semantics. However, we still need to extend our trusted core to a bigger kernel, as the \emph{implementation} has not been formally verified.

Machine integers were also added to the \Coq{} system~\cite{machineintegers}. The usual evaluation inside \Coq{} (\lstinline|compute|) uses a special inductive type for cyclic integers, but the virtual machine uses \OCaml's machine integers. This allows for a big speed-up, for which we pay by having to trust (the virtual machine and) that \OCaml{} treats these integers correctly. The time difference between computation with \Coq's int and \OCaml's \lstinline|Big_int| is about a factor of 20~\cite{thesisSpiwack} on primality tests. 

\section{Metric spaces}\label{section:metricspaces}\label{section:completion-monad}
Having completed our brief description of the \Coq-system, we now turn to O'Connor's formalization of exact real numbers~\cite{OConnor:mscs}. Traditionally, a metric space is defined as a set $X$ with a metric function $d:X \times X → \R^+$ satisfying certain axioms. We use a more relaxed definition of a metric space that does not require the metric be a function; see also~\cite{Richman:2008}. The metric is represented via a (respectful) ball relation $\ballsym: \Q_+ → X → X → \prop$ satisfying:
\begin{lstlisting}
  msp_refl : $∀ x\,ε,\ \ball {ε} x x$
  msp_sym : $∀ x\,y\,ε,\ \ball {ε} x y → \ball {ε} y x$
  msp_triangle : $∀ x\,y\,z\,ε_1\,ε_2,\ \ball {ε_1} x y → \ball {ε_2} y z → \ball {ε_1 + ε_2} x z$
  msp_closed : $∀ x\,y\,ε,\ (∀ δ,\ \ball {ε + δ} x y) → \ball {ε} x y$
  msp_eq : $∀ x\,y,\ (∀ ε,\ \ball {ε} x y) → x = y$
\end{lstlisting}
The ball relation $\ball{ε} x y$ expresses that the points $x$ and $y$ are within $ε$ of each other. We call this a ball relationship because the partially applied relation $\ballsym^X_{ε}\,x : X → \prop$ is a predicate that represents the closed ball of radius $ε$ around the point $x$. For example, the ball relation on $\Q$ is $\ball[\Q] {ε} x y := |x - y| ≤ ε$.


We will introduce the completion of a metric space as a monad. In order to do this we will first introduce monads.

\paragraph{Monads.} Moggi~{\cite{moggi:1989}} recognized that many non-standard forms of computation may be modeled by monads\footnote{In category theory one would speak about the Kleisli category of a (strong) monad.}. Wadler~{\cite{Wadler92a}} popularized their use in functional programming. Monads are now an established tool to structure computation with side-effects. For instance, programs with input $X$ and output $Y$ which have access to a mutable state $S$ can be modeled as functions of type $X \times S → Y \times S$, or equivalently $X → (Y \times S)^S$. The type constructor $\monad Y := (Y \times S)^S$ is an example of a monad. Similarly, partial functions may be modeled by maps $X → Y_{\bot}$, where $Y_{\bot} := Y + ()$ is a monad.

The formal definition of a (strong) monad is a triple $(\monad, \return, \bind)$ consisting of a type
constructor $\monad$ and two functions:
\begin{flalign*}
   \return &: X → \monad X \\
   \bind   &: (X → \monad Y) → \monad X → \monad Y
\end{flalign*}
We will denote $\return x$ as $\returnShort x$, and $\bind f$ as $\bindShort f$. These two operations must satisfy:
\begin{flalign*}
  \bind\ \return\ a &= a \\
  \bindShort f\,  \returnShort a  &= f\, a \\
  \bindShort f\, (\bindShort g\, a) &= \bind\ (\bindShort f ∘ g)\, a
\end{flalign*}
\paragraph{Completion monad.}
The completion of a metric space $X$ is defined by:
\[ 
	\complete X := \{ f : \Q_+ → X \separator ∀ ε_1\,ε_2,\ \ball {ε_1 + ε_2} {(f\,ε_1)} {(f\,ε_2)} \}.
\]
Given metric spaces $X$ and $Y$, a function $f : X → Y$ is \emph{uniformly continuous} with \emph{modulus} $μ_f : Q_+ → Q_+$ if:
\[
	∀ ε\,x_1\,x_2,\ \ball {μ_f ε} {x_1} {x_2} → \ball {ε} {(f\,x_1)} {(f\,x_2)}.
\]
Completion is a monad on the category of metric spaces with uniformly continuous functions. The function $\return : X → \complete X$ defined by $λ x\, ε,\,x$ is the embedding of a metric space in its completion. Moreover, a uniformly continuous function $f : X → \complete Y$ with modulus $μ_f$ can be lifted to operate on complete metric spaces as $\bind f : \complete X → \complete Y$ defined by
$λ x\,ε,\, f\, (x\, (µ_f \frac{ε}{2}))\, \frac{ε}{2}$. In fact, the text above contains a white lie: we need a minor restriction to prelength spaces~\cite{Oconnor:real}.

One advantage of this approach is that it helps us to work with simple representations. Let $\R := \complete \Q$. Then to specify a function from $\R → \R$, we define a uniformly continuous function $f : \Q → \R$, and obtain $\bindShort f : \R → \R$ as the required function. Hence, the completion monad allows us to do in a structured way what was already folklore in constructive mathematics: to work with simple, often decidable, approximations to continuous objects.

\section{Abstract interfaces using type classes}
\label{section:interfaces}
An important part of this work is to further develop the algebraic hierarchy based on type classes by Spitters and van der Weegen~\cite{math-classes}. Especially, we have formalized some order theory and developed interfaces for mathematical operations common in programming languages such as shift and power. This layer of abstraction makes both proof engineering and programming more flexible: it avoids duplication of code, it introduces a canonical way to refer to operations and properties, both by names and notations, and it allows us to easily swap different implementations of number representations and their operations. First we will briefly recap the design decisions made in~\cite{math-classes}.

Algebraic structures are expressed in terms of a number of carrier sets, a number of relations and operations, and a number of laws that the operations satisfy. One way of describing such a structure is by a \emph{bundled representation}: one uses a dependently typed record that contains the carrier, operations and laws. For example a semigroup can be represented as follows. (The fields \lstinline|sg_car| and \lstinline|sg_proper| support our explicit handling of naive set theory in type theory.)
\label{lstlisting:semigroup_bundled}
\begin{lstlisting}
Record SemiGroup : Type := { 
  sg_car :> Setoid ;
  sg_op : sg_car → sg_car → sg_car ;
  sg_proper : Proper ((=) ==> (=) ==> (=)) sg_op ;
  sg_ass : ∀ x y z, sg_op x (sg_op y z) = sg_op (sg_op x y) z) }
\end{lstlisting}  
However, this approach has some serious limitations, the most important one being a lack of support for \emph{sharing} components. For example, suppose we group together two \lstinline|CommutativeMonoid|s in order to create a \lstinline|SemiRing|. Now awkward hacks are necessary to establish equality between the carriers. A second problem is that if we stack up these records to represent higher structures the projection paths become increasingly long.

Historically these problems have been an acceptable trade-off because \emph{unbundled representations}, in which the carrier and operations are parameterized, introduce even more problems. 
\begin{lstlisting}
Record SemiGroup {A} (e : A → A → Prop) (sg_op : A → A → A) : Prop := { 
  sg_proper : Proper (e ==> e ==> e) sg_op ;
  sg_ass : ∀ x y z, e (sg_op x (sg_op y z)) (sg_op (sg_op x y) z) }
\end{lstlisting}
There is nothing to bind notation to, no structure inference, and declaring and passing requires too much manual bookkeeping. Spitters and van der Weegen have proposed a use of \Coq's new type class machinery that resolves many of the problems of unbundled representations. Our current experiment confirms that this is a viable approach. 

An alternative solution is provided by packed classes~\cite{packed} which use an alternative, and older, implementation of a semblance of type classes: canonical structures. Yet another approach would use modules. However, as these are not fist class, we would be unable to define, e.g.\ homomorphisms between algebraic structures.

An \emph{operational type class} is defined for each operation and relation.
\begin{lstlisting}
Class Equiv A := equiv: relation A.
Infix "=" := equiv: type_scope.
Class RingPlus A := ring_plus: A → A → A.
Infix "+" := ring_plus.
\end{lstlisting}
Now an algebraic structure is just a type class living in \prop{} that is parametrized by its carrier, relations and operations. This class contains all laws that the operations should satisfy. Since the operations are unbundled we can easily support sharing. For example let us consider the \lstinline|SemiRing| interface.
\begin{lstlisting}
Class SemiRing A {e : Equiv A} {plus: RingPlus A} 
   {mult: RingMult A} {zero: RingZero A} {one: RingOne A} : Prop := { 
  semiring_mult_monoid :> @CommutativeMonoid A e mult one ;
  semiring_plus_monoid :> @CommutativeMonoid A e plus zero ;
  semiring_distr :> Distribute (.*.) (+) ;
  semiring_left_absorb :> LeftAbsorb (.*.) 0 }.
\end{lstlisting}   
Without type classes it would be a burden to manually carry around the carrier, relations and operations. However, because these parameters are just type class instances, the type class machinery will perform that job for us. For example,
\begin{lstlisting}
Lemma example `{SemiRing R} x : 1 * x = x + 0.
\end{lstlisting}
The backtick instructs \Coq{} to automatically insert implicit declarations, namely \lstinline|e plus mult zero one|. It further lets us omit a name for the \lstinline|SemiRing R| parameter itself as well. All of these parameters will be given automatically generated names that we will never refer to. Furthermore, instance resolution will automatically find instances of the operational type classes for the written notations. Thus the above is really:
\begin{lstlisting}
Lemma example {R e plus mult zero one} {P : @SemiRing R e plus mult zero one} x : 
  @equiv R e 
    (@ring_mult R mult (@ring_one R one) x) 
    (@ring_plus R plus x (@ring_zero R zero)).
\end{lstlisting}

The syntax \lstinline|:>| in the definition of \lstinline|SemiRing| declares certain fields as substructures. That means, a \lstinline|SemiRing| can be seen as a \lstinline|CommutativeMonoid| and each time a \lstinline|CommutativeMonoid| instance is needed, a \lstinline|SemiRing| can be used instead. This syntax should not be confused with the similar syntax for coercions in records (e.g. in the bundled representation of a \lstinline|SemiGroup| on p.~\pageref{lstlisting:semigroup_bundled}).

This approach to interfaces proved useful to formalize a standard algebraic hierarchy. Combined with category theory and universal algebra, $\N$ and $\Z$ are represented as interfaces specifying an initial \lstinline|SemiRing| and initial \lstinline|Ring|~\cite{math-classes}. These abstract interfaces for the naturals and integers make it easier to change the concrete representation in the future.
No such simple specification for $\Q$ seems to exists, so we choose to specify it as the field of fractions of \Z. More precisely, $\Q$ is specified as a \lstinline|Field| containing $\Z$ that moreover can be embedded into the field of fractions of $\Z$.
\begin{lstlisting}
Inductive Frac R `{e : Equiv R} `{zero : RingZero R} : Type := 
  frac { num : R ; den : R ; den_nonzero : den ≠ 0 }.
Class RationalsToFrac (A : Type) := rationals_to_frac : ∀ B `{Integers B}, A → Frac B.
Class Rationals A {e plus mult zero one opp inv} `{U : !RationalsToFrac A} : Prop :=  { 
  rationals_field :> @Field A e plus mult zero one opp inv ; 
  rationals_frac :> ∀ `{Integers Z}, Injective (rationals_to_frac A Z) ; 
  rationals_frac_mor :> ∀ `{Integers Z}, SemiRing_Morphism (rationals_to_frac A Z) ; 
  rationals_embed_ints :> ∀ `{Integers Z}, Injective (integers_to_ring Z A) }.
\end{lstlisting}
\vspace*{-3mm}
\subsection{Order theory}
To abstract from $\N$, $\Z$, $\Q$ and \R\ and their various implementations, we provide a basic library for ordered algebraic structures. For example,
\begin{lstlisting}
Class RingOrder `{Equiv A} `{RingPlus A} `{RingMult A} `{RingZero A} 
	(o : Order A) := { 
  ringorder_partialorder :> PartialOrder ($≤$) ;
  ringorder_plus :> `(OrderPreserving (z +));
  ringorder_mult : `(0 ≤ x → ∀ y, 0 ≤ y → 0 ≤ x * y) }.
\end{lstlisting}
To apply this to $\N$, which is merely a semiring, we introduce the, apparently new, notion of a \lstinline|SemiRingOrder|. Every \lstinline|RingOrder| is a \lstinline|SemiRingOrder|.
\begin{lstlisting}
Class SemiRingOrder `{Equiv A} `{RingPlus A} `{RingMult A} `{RingZero A} 
	(o : Order A) := { 
  srorder_partialorder :> PartialOrder ($≤$) ;
  srorder_plus : `(x ≤ y ↔ ∃ z, 0 ≤ z ∧ y = x + z) ;
  srorder_mult : `(0 ≤ x → ∀ y, 0 ≤ y → 0 ≤ x * y) }.
\end{lstlisting}
This allows us to refer by canonical names to lemmas as those shown below for $\N$, $\Z$, $\Q$ and the dyadics.
\begin{lstlisting}
Lemma plus_compat x1 y1 x2 y2 : x1 ≤ y1 → x2 ≤ y2 → x1 + x2 ≤ y1 + y2.
Lemma sprecedes_1_2 : 1 < 2.
\end{lstlisting}
For instances of $\N$, $\Z$, $\Q$ it is easy to define an order satisfying these interfaces:
\begin{lstlisting}
Instance nat_precedes `{Naturals N} : Order N | 10 :=  λ x y, ∃ z, y = x + z.
\end{lstlisting}
However, often we encounter an a priori different order on a structure, most likely an order defined in \Coq's standard library (like \lstinline|Nle| on \lstinline|N|). Therefore we prove that an arbitrary order satisfying these interfaces while also being a \lstinline|TotalOrder| uniquely specifies the order on $\N$, $\Z$ and $\Q$. For example:
\begin{lstlisting}
Context `{Naturals N} `{Naturals N2} {f : N → N2} `{!SemiRing_Morphism f}
  {o1 : Order N} `{!SemiRingOrder o1} `{!TotalOrder o1}
  {o2 : Order N2} `{!SemiRingOrder o2} `{!TotalOrder o2}.
Global Instance: OrderEmbedding f.
\end{lstlisting}
Unfortunately \Coq{} has no support to have an argument be `inferred if possible, generalized otherwise'; see~\cite{math-classes}. When declaring a parameter of \lstinline|RingOrder|, one is often in a context where most of its components are already available. Usually, only the parameter \lstinline|Order| has to be introduced. The current workaround in these cases involves providing names for components that are then never referred to, which is a bit awkward. In the above it would much nicer to write:
\begin{lstlisting}
Context `{Naturals N} `{Naturals N2} {f : N → N2} `{!SemiRing_Morphism f}
  `{!SemiRingOrder N} `{!TotalOrder N} `{!SemiRingOrder N2} `{!TotalOrder N2}.
Global Instance: OrderEmbedding f.
\end{lstlisting}

\subsection{Basic operations}
The operation \lstinline|nat_pow| is most commonly, but inefficiently, defined as repeated multiplication and the operation \lstinline|shiftl| is defined as repeated multiplication by 2. Instead we specify the desired behavior of these operations. This approach allows for different implementations for different number representations and avoids definitions and proofs  becoming implementation dependent.

We introduce interfaces that specify the behavior of the operations \lstinline|abs|, \mbox{\lstinline|shiftl|,} \lstinline|nat_pow| and \lstinline|int_pow|. Again there are various ways of specifying these interfaces: with $\Sigma$-types, bundled or unbundled. In general, \mbox{$\Sigma$-types} are convenient for functions whose specification is easy, for example:
\begin{lstlisting}
Class Abs A `{Equiv A} `{Order A} `{RingZero A} `{GroupInv A} 
	:= abs_sig: ∀ (x : A), { y : A | (0 ≤ x → y = x) ∧ (x ≤ 0 → y = -x)}.
Definition abs `{Abs A} := λ x : A, ` (abs_sig x).
\end{lstlisting} 
However, for more complex operations, such as \lstinline|shiftl|, such an interface is different from the usual mathematical specification because we cannot quantify over all possible input values. Now there are two ways: a bundled or an unbundled interface. Since these interfaces are not used for hierarchies the disadvantages of the latter do not apply. Let us first describe the former approach.
\begin{lstlisting}
Class ShiftL A B `{Equiv A} `{Equiv B} `{RingOne A} 
	`{RingPlus A} `{RingMult A} `{RingZero B} `{RingOne B} `{RingPlus B} := {
  shiftl : A → B → A ;
  shiftl_proper : Proper ((=) ==> (=) ==> (=)) shiftl ;
  shiftl_0 :> RightIdentity shiftl 0 ;
  shiftl_S : ∀ x n, shiftl x (1 + n) = 2 * shiftl x n }.
Infix "≪" := shiftl (at level 33, left associativity).
\end{lstlisting}
Although this interface seems reasonable, it does not work well in \Coq. The \lstinline|simpl| tactic which is used to simplify a goal will unfold occurrences of \lstinline|shiftl| to their underlying definition (for example in case of \lstinline|BigN|, the expression \lstinline|x ≪ n| becomes \lstinline|BigN.shiftl x n|). This is rather inconvenient because \Coq{} will then be unable to use lemmas concerning $≪$ for rewriting. This problem is caused because \lstinline|shiftl| is a projection of a record, which is in fact an $ι$-redex (reduction of pattern-matching over a constructed term) that will be unfolded by \lstinline|simpl|. Currently there seems to be no way to adjust the behavior of \lstinline|simpl| to remove this inconvenience. A similar problem was already observed in \Ssreflect~\cite{ssr}.

Instead we use an unbundled interface, which has a lot in common with our interfaces for algebraic structures. Now \lstinline|shiftl| no longer contains an $ι$-redex. 
\begin{lstlisting}
Class ShiftL A B := shiftl: A → B → A.
Infix "≪" := shiftl (at level 33, left associativity).
Class ShiftLSpec A B (sl : ShiftL A B) `{Equiv A} `{Equiv B} `{RingOne A} 
	`{RingPlus A} `{RingMult A} `{RingZero B} `{RingOne B} `{RingPlus B} := {
  shiftl_proper : Proper ((=) ==> (=) ==> (=)) (≪) ;
  shiftl_0 :> RightIdentity (≪) 0 ;
  shiftl_S : ∀ x n, x ≪ (1 + n) = 2 * x ≪ n }.
\end{lstlisting}
We do not specify \lstinline|shiftl| as \lstinline|shiftl x n = x * 2 ^ n| since on the dyadics we cannot take a negative power while we can shift by a negative integer.

\subsection{Decision procedures}
\label{section:decision}
The \lstinline|Decision| type class collects types with a decidable equality~\cite{math-classes}.
\begin{lstlisting}
Class Decision P := decide: sumbool P (¬P).
\end{lstlisting}
Using this type class we can declare a parameter \lstinline|`{∀ x y, Decision (x ≤ y)}| to describe a decider for $≤$ and say \lstinline|decide (x ≤ y)| to decide whether \lstinline|x ≤ y| or not. This type class allows us to easily define additional deciders, like the one for the strict order. We have to be careful however. Consider the order on the dyadics.
\begin{lstlisting}
Global Instance dy_precedes: Order Dyadic := λ (x y : Dyadic), 
  ZtoQ (mant x) * 2 ^ (expo x) ≤ ZtoQ (mant y) * 2 ^ (expo y)
\end{lstlisting}
Now, \lstinline|decide (x ≤ y)| is actually \lstinline|@decide Dyadic (x ≤ y)$\ $dyadic_dec|, where \lstinline|dyadic_dec| is the computational conclusion of the decision. Due to eager evaluation, and the absence of dead code removal, the second argument, \lstinline|x ≤ y|, is also evaluated. Evaluation of this argument results in a conversion of \lstinline|x| and \lstinline|y| into \lstinline|Q|, as described above. But since this argument is just a proposition it is later thrown away. We avoid this problem introducing a λ-abstraction.
\begin{lstlisting}
Definition decide_rel `(R : relation A) {dec : ∀ x y, Decision (R x y)} 
	(x y : A) : Decision (R x y) := dec x y.
\end{lstlisting}
We can now define:
\begin{lstlisting}
Context `{!PartialOrder (≤)} {!TotalOrder (≤)} `{∀x y, Decision (x ≤ y)}.
Global Program Instance sprecedes_dec: ∀ x y, Decision (x < y) | 9 := λ x y,
  match decide_rel (≤) y x with
  | left E => right _
  | right E => left _
  end.
\end{lstlisting}

\subsection{Approximate rationals}\label{section:approx_rationals}
To make our implementation of the reals independent of the underlying dense set, we provide an abstract specification of \emph{approximate rationals} inspired by the notion of \emph{approximate fields} which is used in the \Haskell\ implementation of the exact reals by Bauer and Kavler~\cite{BauerKavkler}. We provide an implementation of this interface by dyadics based on \Coq's machine integers.

Our interface describes an ordered ring containing \lstinline|Z| that is dense in \lstinline|Q|. Here \lstinline|Z| are the binary integers from \Coq's standard library, and \lstinline|Q| are the rationals based on these binary integers. We do not parametrize by arbitrary integer and rational implementations because they are hardly used for computation.

Also, for efficient computation, this interface contains the operations: approximate division, normalization, an embedding of \lstinline|Z|, absolute value, power by \lstinline|N|, shift by \lstinline|Z|, and decision procedures for both equality and order.
\begin{lstlisting}
Class AppDiv AQ := app_div : AQ → AQ → Z → AQ.
Class AppApprox AQ := app_approx : AQ → Z → AQ.
Class AppRationals AQ {e plus mult zero one inv} `{!Order AQ} 
	{AQtoQ : Coerce AQ Q_as_MetricSpace} `{!AppInverse AQtoQ} 
	{ZtoAQ : Coerce Z AQ} `{!AppDiv AQ} `{!AppApprox AQ} 
   `{!Abs AQ} `{!Pow AQ N} `{!ShiftL AQ Z} 
   `{∀x y : AQ, Decision (x = y)} `{∀x y : AQ, Decision (x ≤ y)} : Prop := {
  aq_ring :> @Ring AQ e plus mult zero one inv ;
  aq_order_embed :> OrderEmbedding AQtoQ ;
  aq_ring_morphism :> SemiRing_Morphism AQtoQ ;
  aq_dense_embedding :> DenseEmbedding AQtoQ ;
  aq_div : ∀ x y k, $\ballsym_{2 ^ k}$('app_div x y k) ('x / 'y) ;
  aq_approx : ∀ x k, $\ballsym_{2 ^ k}$('app_approx x k) ('x) ;
  aq_shift :> ShiftLSpec AQ Z (≪) ;
  aq_nat_pow :> NatPowSpec AQ N (^) ;
  aq_ints_mor :> SemiRing_Morphism ZtoAQ }.
\end{lstlisting}
O'Connor~\cite{OConnor:mscs} keeps the size of the rational numbers small to avoid efficiency problems. He introduces a function \mbox{\lstinline|approx x ε|} that yields the `simplest' rational number between \lstinline|x - ε| and \lstinline|x + ε|. In our interface we modify the \lstinline|approx| function slightly: \lstinline|app_approx x k| yields an arbitrary element between \lstinline|x - $2^k$| and \mbox{\lstinline|x + $2^k$|.}
Using this function we define the compress operation on the real numbers: \mbox{\lstinline|compress := bind (λ ε, app_approx x (Qdlog2 ε))|} such that \lstinline|compress x = x|.

In Section~\ref{section:series} we will explain our choice of using a power of 2 to specify the precision of \lstinline|app_div| and \lstinline|app_approx|. In the remainder of this section we briefly describe our implementation by the dyadics.

The dyadic rationals are numbers of the shape $n * 2 ^ e$ for $n,e \in \Z$. In order to remain independent of an integers implementation, we abstract over it. For our eventual implementation of the approximate rationals we use \Coq's machine integers, \lstinline|bigZ|. Now given an arbitrary integer implementation \lstinline|Int| it is straightforward to define the dyadics. Here we will just show the ring operations.
\begin{lstlisting}[mathescape=false,breaklines=false]
Notation "x↾p" := (exist _ x p) (at level 20).
Record Dyadic := dyadic { mant : Int ; expo : Int }.
Infix "$" := dyadic (at level 80).
Global Instance dy_inject: Coerce Int Dyadic := λ x, x $ 0.
Global Instance dy_opp: GroupInv Dyadic := λ x, -mant x $ expo x.
Global Instance dy_mult: RingMult Dyadic := λ x y, mant x * mant y $ expo x + expo y.
Global Instance dy_0: RingZero Dyadic := ('0:Dyadic).
Global Instance dy_1: RingOne Dyadic := ('1:Dyadic).
Global Program Instance dy_plus: RingPlus Dyadic := λ x y, 
  if decide_rel (≤) (expo x) (expo y)
  then mant x + mant y ≪ (expo y - expo x)↾_ $ min (expo x) (expo y)
  else mant x ≪ (expo x - expo y)↾_ + mant y $ min (expo x) (expo y).
\end{lstlisting}
In this code \lstinline|shiftl| has type \lstinline|Int → Int⁺ → Int|, where \lstinline|Int⁺| is a $\Sigma$-type describing the non-negative elements of \lstinline|Int|. Therefore, in the definition of \lstinline|dy_plus| we have to equip \lstinline|expo y - expo x| with a proof that it is in fact non-negative.

\section{Power series}\label{section:series}
Elementary transcendental functions as \lstinline|exp|, \lstinline|sin|, \lstinline|ln| and \lstinline|arctan| can be defined by their power series. If the coefficients of a power series are alternating, decreasing and have limit 0, then we obtain a fast converging sequence with an easy termination proof. For $-1 ≤ x ≤ 0$,
\[
	\exp\ x = \sum_{i=0}^\infty \frac {x^i} {i!}
\]
is of this form. To approximate $\exp\ x$ with error $ε$ we take the partial sum until $\frac{x^i}{i!} ≤ ε$.
In order to implement this efficiently we use a stream representing the series and define a function that sums the required number of elements. For example, the series \lstinline|1, a, a$^2$, a$^3$, ...| is defined by the following stream.
\begin{lstlisting}
CoFixpoint powers_help (c : A) : Stream A := Cons c (powers_help (c * a)).
Definition powers : Stream A := powers_help 1.
\end{lstlisting}
Streams in \Coq, like lists in \Haskell, are lazy. So, in the example the multiplications are accumulated.

Since \Coq{} only allows structural recursion (and guarded co-recursion) it requires some work to convince \Coq{} that our algorithm terminates. Intuitively, one would describe the limit as an upperbound of the required number of elements using the \lstinline|Exists| predicate.
\begin{lstlisting}
Inductive Exists A (P : Stream A → Prop) (x : Stream) : Prop :=
  | Here : P x → Exists P x
  | Further : Exists P (tl x) → Exists P x.
\end{lstlisting}
This approach leads to performance problems. The upperbound, encoded in unary format, may become very large while generally only a few terms are necessary. Due to \lstinline|vm_compute|'s eager evaluation scheme, this unary number will be computed before summing the series. Instead we use \lstinline|LazyExists|~\cite{oconnor-thesis}.
\begin{lstlisting}
Inductive LazyExists A (P : Stream A → Prop) (x : Stream A) : Prop :=
  | LazyHere : P x → LazyExists P x
  | LazyFurther : (unit → LazyExists P (tl x)) → LazyExists P x.
\end{lstlisting}

O'Connor's \lstinline|InfiniteAlternatingSum $s$| returns the real number represented by the infinite alternating sum over $s$, where the stream $s$ is decreasing, non-negative and has limit 0.
We have extended this in two ways. First, by generalizing some of the work to abstract structures. Second, as we do not have exact division on approximate rationals, we extended his algorithm to work with approximate division. The latter required changing \lstinline|InfiniteAlternatingSum $s$| to \lstinline|InfiniteAlternatingSum $n\ d$| which computes the infinite alternating sum of the stream $λ i,\frac {n_i}{d_i}$. This allows us to postpone divisions. Also, we have to determine both the length of the partial sum and the required precision of the divisions. To do so we find a $k$ such that:
\begin{equation}
	\ball {\frac{ε}{2}} {(\appdiv\ n_k\ d_k\ (\log \frac{ε}{2k}) + \frac{ε}{2k})} 0.\label{equation:series_ball}
\end{equation}
Now $k$ is the length of the partial sum, and $\frac{ε}{2k}$ is the required precision of division. Using O'Connor's results we have verified that these values are correct and such a $k$ indeed exists for a decreasing, non-negative stream with limit 0.

As noted in Section~\ref{section:approx_rationals}, we have specified the precision of division in powers of 2 instead of using a rational value. This allows us to replace (\ref{equation:series_ball}) with:
\[
	\ball {\frac{ε}{2}} {(\appdiv\ n_k\ d_k\ (\log\ ε - (k + 1)) + 1 ≪ (\log\ ε - (k + 1)))} 0.
\]
Here $k$ is the length of the partial sum, and $2^l$, where $l = \log\ ε - (k + 1)$, is the required precision of division. This variant can be implemented without any arithmetic on the rationals and is thus much more efficient.

This method gives us a fast way to compute the infinite alternating sum, in practice, only a few extra terms have to be computed and due to the approximate division the auxiliary results are kept as small as possible.

Using this method to compute infinite alternating sums we have so far implemented \exp\ and \arctan. Furthermore, we extend the exponential to its complete domain by repeatedly applying the following formula.
\begin{equation}
	\exp\ x = (\exp (x ≪ 1))^2 \label{equation:exp_squaring}
\end{equation}
Our tests have shown that reducing the input to a value between $-2^k ≤ x ≤ 0$ for $50 ≤ k$ yields major performance improvements as the series will converge much faster. For higher precisions setting it to $75 ≤ k$ gives even better results.

By defining $\arctan$ on $[0,1)$, we can define the Machin-like formula
\[
	\pi := 176 * \arctan \frac{1}{57} + 28 * \arctan \frac{1}{239} 
		- 48 * \arctan \frac{1}{682} + 96 * \arctan \frac{1}{12943}.
\]
Since we do not have exact division on the approximate rationals, we see here the purpose of parameterizing infinite sums by two streams.

\section{Square root}\label{section:Wolfram}
We use Wolfram's algorithm~\cite[p.913]{wolfram2002new} for computing the square root. Its complexity is linear, in fact it provides a new binary digit in each step. We aim to investigate Newton iteration in future work.
\begin{lstlisting}
Context `(Pa : 1 ≤ a ≤ 4).
Fixpoint AQroot_loop (n : nat) : AQ * AQ :=
  match n with
  | O => (a, 0)
  | S n =>
     let (r, s) := AQroot_loop n in
     if decide_rel (≤) (s + 1) r
     then ((r - (s + 1)) ≪ (2:Z), (s + 2) ≪ (1:Z))
     else (r ≪ (2:Z), s ≪ (1:Z))
  end.
\end{lstlisting}
Three easy invariants allow us to prove this series converges to the square root.
\begin{lstlisting}
Lemma AQroot_loop_invariant1 (n : nat) :
 snd (AQroot_loop n) * snd (AQroot_loop n) + 4 * fst (AQroot_loop n) = 4 * 4 ^ n * a.

Lemma AQroot_loop_invariant2 (n : nat) :
 fst (AQroot_loop n) ≤ 2 * snd (AQroot_loop n) + 4.
Lemma AQroot_loop_fst_bound (n : nat) :
 fst (AQroot_loop n) ≤ 2 ^ (3 + n).
\end{lstlisting}
 
 
\section{Benchmarks}\label{section:bench}
The first step in this research was to create a \Haskell{} prototype based on O'Connor's implementation of the real numbers in \Haskell~\cite{OConnor:mscs}. The second step was to implement this prototype in \Coq. Currently, our \Coq{} development contains the field operations, computation of power series, \exp, \arctan, $\pi$ and the square root. Apart from the square root, the correctness of these operations has been verified in the \Coq{} system.

In this section we present some benchmarks comparing the old and the new implementation, both in \Haskell{} and \Coq. All benchmarks have been carried out on an Intel Core Quad 2.4 GHz with 8GB of memory running \DebianFull{} with kernel 2.6.32. The sources of our developments can be found at \url{http://robbertkrebbers.nl/research/reals}. 

\begin{table}[tp]
\centering
\begin{tabular}{|c|c|c|c|}
\hline
Expression & \quad Decimals \quad  & \quad O'Connor \quad & \quad  Krebbers/Spitters \quad  \\
\hline
$\sin\ (\sin\ (\sin\ 1))$ & 10,000 & 71s & 5s\\
$\cos\ (10 ^{50})$ & 10,000 & 2.7s & 0.6s\\
$\tan\ (\sqrt 2) + \arctanh\ (\sin\ 1)$ & 500 & 133s & 2.2s\\
\hline
\end{tabular}
\caption{\Haskell{}, compiled with \texttt{ghc} version 6.12.1, using \texttt{-O2}.}
\label{table:haskellO2}
\end{table}

\begin{table}[tp]
\centering
\begin{tabular}{|c|c|c|c|}
\hline
Expression & \quad Decimals \quad  & \quad O'Connor \quad & \quad  Krebbers/Spitters \quad  \\
\hline
$\pi$ & 300 & 55s & 0.8s\\
$\exp\ (\exp\ (\exp\ (\frac 1 2 ))) $ & 25 & 123s & 0.23s\\
$\exp\ \pi - \pi$ & 25 & 52s & 0.1s\\
$\arctan\ \pi$ & 25 & 134s & 1.0s\\
\hline
\end{tabular}
\caption{\Coq{} trunk revision 13841.}
\label{table:coq}
\end{table}

Table~\ref{table:haskellO2} shows some benchmarks in \Haskell{} with compiler optimizations enabled (\texttt{-O2}) and Table~\ref{table:coq} compares our \Coq{} implementation with O'Connor's. More extensive benchmarking shows that our \Haskell{} implementation generally benefits from a 15 times speed up while the speed up in \Coq{} is usually more than a 100 times. This difference is explained by the fact that O'Connor's \Haskell{} implementation already used fast integers, while his \Coq{} implementation did not. In the same times as shown in Table~\ref{table:coq} for the old implementation, the new implementation is able to compute the first 2,000 decimals of $\pi$, 450 decimals of $\exp\ (\exp\ (\exp\ (\frac 1 2 ))) $, 425 decimals of $\exp\ \pi - \pi$ and 85 decimals of $\arctan\ \pi$. This is an improvement of up to 18 times of the number of decimals.

It is interesting to notice that $\pi$ and \arctan{} benefit the least from our improvements, as we are unaware of an optimization similar to the squaring trick for \exp{} (Section~\ref{section:series}, Equation~\ref{equation:exp_squaring}).

We conclude this section with a comparison between the performance of Wolfram's algorithm in \Coq{} and \Haskell. The \Haskell{} prototype (without compiler optimizations) is quite fast, computing 10,000 iterations (giving 3,010 decimals) of $\sqrt2$ takes 0.2s. In \Coq{} it takes 11.6s using type classes and 11.3s without type classes. Here we exclude the time spend on type class resolution. Thus type classes cause only a 3\% performance penalty on computations.

Unfortunately, the \Coq-implementation is slow compared to \Haskell. Laurent Théry suggested that this is due to the representation of the fast integers, which uses a tree with a fixed depth and when the size of the integer becomes too big uses a less optimal representation. Increasing the size of the tree representation and avoiding an inefficiency in the implementation of shifts reduces this time to 7.5s.

\section{Conclusions and Related work}
We have greatly improved the performance of real number computation in \Coq{} using \Coq's new machine integers. We produced highly structured and abstract code using type classes with no apparent performance penalty. Moreover, \Coq's notation mechanism combined with unicode characters gives nicely readable statements and proofs. Type classes were a great help in our work. However, the current implementation of instance resolution is still experimental and at times too slow (at compile time).

Canonical structures provide an alternative, and partially complementary, implementation of type classes~\cite{adhoc}. By choice, canonical structures restrict to deterministic proof search, this makes them more efficient, but also somewhat more intricate to use. The use of canonical structures by the \Ssreflect{} team~\cite{packed} makes it plausible that with some effort we could have used canonical structures for our work instead. However, the \Ssreflect-library is currently not suited for setoids which are crucial to us. The integration of unification hints~\cite{Hints} into \Coq\ may allow a tighter integration of type classes and canonical structures.

We needed to adapt our correctness proofs to prevent the virtual machine from eagerly evaluating them. Lazy evaluation for \prop\ would have allowed us to use the original proofs.

The experimental \lstinline|native_compute| performs evaluation by compilation to native \OCaml{} code. This approach uses the \OCaml{} compiler available and is interesting for heavy compilation. Our first experiments indicate a 10 times speed up with Wolfram iteration. Unfortunately, \lstinline|native_compute| does not work with \Coq{} trunk yet, so we were unable to test it with our implementation of the reals.

The \Flocq{} project~\cite{BolMel11} formalizes floating-points in \Coq. It provides a library of theorems on a multi-radix multi-precision arithmetic and supports efficient numerical computations inside \Coq. However, the current library is still too limited for our purposes, but in the future it should be possible to show that they form an instance of our approximate rationals. This may allow us to gain some speed by taking advantage of fine grained algorithms on the floats instead of our more straightforward ones.

The encoding of real numbers as streams of `bits' is potentially interesting. However, currently there is a big difference in performance. The computation of 37 decimals of the square root of 1/2 by Newton iteration~\cite{JulienP09}, using the framework described in~\cite{bertot2007affine,julien2008certified}, took 12s. This should be compared with our use of the Wolfram iteration, which gives only linear convergence, but with which we nevertheless obtain 3,000 decimals in in a similar time. On the other hand, the efficiency of $\pi$ in their framework is comparable with ours. Berger~\cite{berger2009coinductive}, too, uses co-induction for exact real computation.

The present work is part of a larger program to use constructive mathematics based on type theory as a programming language for exact analysis. This should culminate in a numerical ODE-solver.

\paragraph{Acknowledgements}
We thank Eelis van der Weegen for many discussions and Pierre Letouzey and Matthieu Sozeau for closing some of our bug reports. We are grateful to the anonymous referees who provided some helpful suggestions.

\bibliographystyle{splncs}
\bibliography{alg}
\end{document}